\documentclass[superscriptaddress,nofootinbib]{revtex4-1}
\usepackage[utf8]{inputenc}
\usepackage{bm}

\usepackage[dvipsnames]{xcolor}
\definecolor{red}{rgb}{0.9, 0,0}
\definecolor{cerulean}{rgb}{0., 0.42,0.9}
\definecolor{navy}{rgb}{0.05, 0.05,0.8}

\usepackage[colorlinks]{hyperref}
\hypersetup{
  colorlinks = true,
  citecolor = red,
	linkcolor = navy
}

\usepackage{bbm}
\usepackage{amsfonts}
\usepackage{mathrsfs}
\usepackage{bbold}
\usepackage{amsmath,amssymb}
\usepackage{subcaption}
\usepackage{siunitx}
\usepackage{slashed}

\usepackage{epsfig}
\usepackage{graphicx}   
\usepackage{url}
\usepackage{float}
\usepackage{soul}
\usepackage{pstricks}
\usepackage{color}
\usepackage{multirow}

\newcommand{\be}{\begin{equation}}
\newcommand{\ee}{\end{equation}}
\newcommand{\bea}{\begin{eqnarray}}
\newcommand{\eea}{\end{eqnarray}}
\newcommand{\beq}{\begin{eqnarray}}
\newcommand{\eeq}{\end{eqnarray}}
\def\bit{\begin{itemize}}
\def\eit{\end{itemize}}
\def\ben{\begin{enumerate}}
\def\een{\end{enumerate}}

\newcommand\DN[1][\relax]{%
\ifx\relax#1\relax\else{}^{#1}\fi \!X}

\DeclareMathAlphabet\mathbfcal{OMS}{cmsy}{b}{n}

\definecolor{cerulean}{rgb}{0., 0.52,0.65}

\graphicspath{{plots/}}

\begin{document}
\title{A Composite Solution to the Neutron Bottle Anomaly}
\author{Surjeet Rajendran}
\affiliation{Department of Physics and Astronomy, The Johns Hopkins University, Baltimore, MD 21218, USA}
\author{Harikrishnan Ramani}
\affiliation{Berkeley Center for Theoretical Physics, Department of Physics,\\ University of California, Berkeley, CA 94720}
\affiliation{Theoretical Physics Group, Lawrence Berkeley National Laboratory, Berkeley, CA 94720}
\begin{abstract}
Measurements of the lifetime of neutrons trapped in a bottle have been consistently shorter than the lifetime measured in neutron beam experiments. With trapping potentials as low as 50 neV and neutron detectors located only at the top of the bottle, this discrepancy could be the result of the soft scattering of dark matter with neutrons. However, it is challenging to obtain the observed loss rate in conventional models of dark matter scattering. We show that this phenomenology is possible in composite models of dark matter where the soft scattering is from dark matter that has been captured and accumulated in the earth. This solution can be tested by placing more neutron detectors around the trap, providing better angular coverage. The phenomenology of soft scattering by trapped composite dark matter is generic and suggests new experimental directions that could be pursued to detect this large class of models. 
\end{abstract}
\maketitle

\section{Introduction}
The lifetime of the neutron, a number of fundamental importance to big bang nucleosynthesis, has recently become a topic of contention. A persistent discrepancy at the level of 3.9 $\sigma$ has been reported between two different techniques used to measure this lifetime. The first technique measures the lifetime by looking at the decay of free neutrons in a beam. In this method, the decay is identified by the presence of charged particles produced as a result of the decaying neutron. Based on the rate of appearance of charged particles, this measurement yields a lifetime of $887.2 \pm 2.2$ s~\cite{yue2013improved}. In the second method, the lifetime is measured by counting the surviving number of trapped ultra-cold neutrons, yielding a lifetime of $877.7 \pm 0.7$ s~\cite{Pattie:2017vsj}, shorter than the measured lifetime in the beam experiments by 9.2 s. These results suggest that neutrons are being lost from their trap without the production of charged particles. 

An enticing avenue to explain this anomaly is to consider the possibility of new interactions between neutrons and the dark sector. A well explored option is to consider the possibility of the neutron decaying invisibly into dark sector particles~\cite{Fornal:2019olr}. While possible, this option is heavily constrained by the stability of nuclei, requiring the existence of a dark particle that is nearly degenerate with the neutron. The existence of such states may also potentially destabilize neutron stars~\cite{McKeen:2018xwc} and thus require more exotic nuclear equations of state. Stringent limits on such models also arise from the stability of hydrogen atoms \cite{McKeen:2020zni}. Given these theoretical challenges, it is interesting to explore other phenomenological avenues to explain this anomaly. 

In this paper, we explore the possibility that the loss of ultra-cold neutrons from the trap is due to scattering between neutrons and dark matter. This is a plausible direction since these ultra-cold neutrons are trapped by a confining potential $\lessapprox$ 50 neV~\cite{Pattie:2017vsj}, making it possible for even soft collisions to cause loss from the trap. The loss of neutrons through energy exchange between them and the environment is a significant systematic in the trapped neutron experiments and there are thus experimental checks on this possibility~\cite{Pattie:2017vsj}. But, these checks are aimed at constraining two loss mechanisms. The first involves collisions between neutrons and background gas. Experimentally, this possibility is constrained by ensuring that the quality of the vacuum in the trap is sufficiently good. The second involves exchange of energy between the neutrons and the magnetic fields used to confine them. In this case, the neutron gains energy slowly and thus rises slowly up the trap. A neutron detector that is sensitive to soft neutrons is placed at the top of the trap and the activity in this detector is used to constrain loss in this channel. But, if the neutron undergoes a scattering event that can give it a sudden kick $\gtrapprox 50$ neV along any direction, this cross-check is inefficient due to the small solid angle subtended by this detector in the trap \cite{ASaunders}. Since the experiments allow for the scattering of dark matter with the neutrons to explain this discrepancy, it is interesting to ask what class of dark matter models could accomplish this phenomenology while being otherwise unconstrained by other experiments. 

It is challenging to explain this anomaly through the scattering of dark matter with neutrons. First, in order to efficiently kick the neutrons, the dark matter has to be reasonably massive. Second, there needs to be enough of this dark matter in the trap in order to account for the observed rate. It can be verified that these requirements cannot be satisfied with the galactic dark matter population. This problem can be overcome if we instead consider the dark matter population that is captured and bound to the earth. This captured dark matter can have significant over-densities \cite{Pospelov:2019vuf}. While slower than the galactic dark matter, this population is nevertheless hot enough to deposit the $\approx$ 50 neV of energy necessary to kick the neutrons out of the trap. The captured dark matter needs to have a mass in the $\approx$ GeV range to have a significant abundance on the surface of the earth: lighter dark matter evaporates, while heavier ones might sink to the core. Furthermore, even in the GeV range, despite the large abundances, dark matter mediator-couplings larger than unity are required to create a large enough neutron disappearance rate that would explain the bottle-beam discrepancy. 

We show that the required dark matter density and cross-section that is consistent with all observational constraints can be realized in models of composite dark matter. Specifically, we consider dark matter blobs consisting of a large number of dark partons which results in a large dark charge under a long range force. Low momentum scattering between this blob and the standard model via this long range ($\sim$ micron) fifth force is coherently enhanced. This enhanced cross-section permits the blobs to be captured in the Earth. Due to the large self-scattering between blobs, a small initial density of captured blobs can seed the capture of additional blobs, resulting in rapid growth of blob density. In the range of parameters allowed by current experimental constraints, this growth is only possible when there is a distribution of blob-masses {\it i.e.} we will consider a small number of heavier blobs that stop and sink deeper into the Earth. Lighter blobs will scatter off these heavier blobs and get captured. Note that such a distribution in blob masses is to be expected in composite dark matter scenarios, since composite systems generically produce such a distribution during ``blob-nucleosynthesis''. These blobs can then undergo the soft scattering needed to explain the lifetime of the trapped neutrons.  Experiments that probe terrestrial dark matter at larger momentum transfer such as cryogenics \cite{Neufeld:2019xes}, dilution refrigerators \cite{private} and metastable isomers \cite{Lehnert:2019tuw} are suppressed due to the momentum dependent form factor for long range interactions or due to loss of coherence. With large cross-sections the thermalized dark matter or atmospheric overburden slow incoming dark matter down to energies undetectable at even surface experiments like SENSEI~\cite{Crisler:2018gci} or CRESST~\cite{angloher2017results}. 

The rest of this paper is organized as follows: in section \ref{sec:model}, we describe a simple model of the blob and in section \ref{sec:capture} describe its captured population on the Earth. In section \ref{sec:bottle}, we evaluate the parameters necessary to explain the lifetime of the trapped neutrons and in section \ref{sec:limits} we discuss existing experimental limits on this parameter space. Finally, we conclude in section \ref{sec:conclusions} where we discuss future experimental prospects to constrain both this particular explanation for the trapped neutron lifetime and comment more generally on how these kinds of dark matter phenomenology can be probed.

\section{Model}
\label{sec:model}

We consider dark matter $f$ with mass $m_{f}$ that is part of a strongly coupled sector that confines at a scale $\Lambda_f \approx m_f$. The $f$
 particles are gauged under a vector $A$ with mass $m_A$ with gauge coupling $g_f$. The neutron (or proton) is assumed to have a ``dark electric dipole moment" with this vector.
 \begin{equation}
   \mathcal{L} \supset \frac{1}{\Lambda} \bar{n} \sigma_{\mu \nu} \gamma_5 F_A^{\mu \nu} {n}+\bar{f}(m_f+D^\mu_A \gamma_\mu) f+m_A^2 A^2
 \end{equation}
We assume $\Lambda \sim 3\times 10^8 \textrm{GeV}$ to be consistent with Supernova constraints \cite{Chang:2018rso,Ramani:2019jam}. 

 At temperatures below $\Lambda_f$, we assume that the dark sector forms blobs with charge $g_{\rm blob}=N_f g_f$ and mass $m_{\rm blob}=N_f m_f$. We make the simplifying assumption that all of $f$-type dark matter is in blobs. We do not need all of the dark matter to be in these blobs - the fraction that is in blobs is denoted by $f_{\rm blob}$. 

For fermions, stability of these blobs requires \cite{Grabowska:2018lnd}
\begin{align}
  N_f &\lesssim \frac{1}{g_f^3}\implies N_f g_f = g_{\rm blob} \lesssim \frac{1}{g_f^2}\nonumber \\
  \implies g_{\rm blob}^3 &\lesssim N_f^2 \implies N_f \gtrsim g_{\rm blob}^\frac{3}{2}
 \end{align}

Let us assume $N_f = \chi g_{\rm blob}^\frac{3}{2}$. When $\chi \gtrapprox 1$, the blob satisfies the above constraint. But, this is not a strict constraint and it can be overcome by further model building, for instance, using a tuned long-range attractive force that cancels against the negative potential energy due to the vector $A$. For this reason, we take the above bound to be a tuning line rather than a strict bound and will consider situations where $\chi \lessapprox 1$, where the model is tuned. Now, 

\begin{equation}
  \Lambda_f\approx m_f = \frac{m_{\rm blob}}{N_f}\approx \frac{m_{\rm blob}}{\chi g_{\rm blob}^\frac{3}{2}}
\end{equation} 
The size of this blob is
\begin{equation}
  R_{\rm blob}=\frac{N_f^\frac{1}{3} }{\Lambda_D}=\frac{\chi^\frac{4}{3}g_{\rm blob}^2}{m_{\rm blob}}
\end{equation}
This gives a maximum momentum transfer, 
\begin{equation}
q_{\rm max}=\frac{m_{\rm blob}}{\chi^\frac{4}{3}g_{\rm blob}^2}
\label{maxqstab}
\end{equation}
 above which coherent enhancement over the whole blob is lost.

\section{Capture}
\label{sec:capture}
The dark matter blobs fall on Earth and a fraction gets captured if it slows down to velocities below the escape velocity. At large charge $g_{\rm blob} \ge 10^9$ this could be due to scattering with rock, and at smaller charge it needs to be aided by a secondary population of blobs with larger mass (and charge). Additional detail for capture is presented in Appendix.\ref{seccap}. For the remainder of this work, we assume that $100\%$ of in-falling dark matter is captured over the Earth's history. The average dark matter density on Earth is, 

\begin{equation}
 \langle n_{\rm DM}^{\rm terr} \rangle =\frac{\pi r_E^2}{\frac{4}{3}\pi r_E^3} n_{\rm DM}^{\rm vir} v_{\rm vir} \approx\frac{10^{15}}{{\rm cm}^3} \frac{T_E}{10^{10}{\rm year}} f_{\rm blob} \frac{\rm GeV}{m_{\rm blob}}
 \label{capturest}
\end{equation}

Here $r_E\approx 6400 \textrm{km}$ is the radius of the earth, $n_{\rm DM}^{\rm vir}\approx\frac{0.3}{\textrm{cm}^3} f_{\rm blob} \frac{ \textrm{GeV}}{m_{\rm blob}}$ is number density of the virial population, $v_{\rm vir} \approx 300 \textrm{km/sec}$ is the dark matter virial velocity and $T_E\approx 10^{10}~\textrm{year}$ is the age of the Earth. 

\subsection{Thermalized GeV DM population}
For the cross-sections we consider, capture initially occurs all over the Earth and is immediately followed by thermalization; this initial density is expected to be uniform over the volume of the Earth. Subsequently in-falling dark matter gets stopped by the first few layers of terrestrial DM. This thermalized dark matter population has rms velocity $v_{\rm th}\approx \sqrt{T_{\rm room}/m_{\rm blob}}$ with the temperature $T_{\rm room}\approx 300$ K, the room temperature. 
There is however subsequent diffusion, with DM diffusing in time $T_E$ a distance,
\begin{equation}
  R_{\rm diff}=\frac{1}{n_{\rm blob} \sigma_T} \left(n_{\rm blob}\sigma_T v_{\rm th} T_E \right)^\frac{1}{2}\approx 10^5 \textrm{km} \frac{m_A}{10 {\rm eV} }\sqrt{\frac{10^{14}/\textrm{cm}^3}{n_{\rm blob}}}\left(\frac{\rm GeV}{m_{\rm blob}}\right)^\frac{1}{4} \gtrsim R_E
\end{equation}
Thus DM spreads all over the earth. 

Temperature and density variations, as well as gravity can significantly modify the density profile on Earth. This Jeans density $n_{\rm jeans}$, was calculated in \cite{Neufeld:2018slx} and it was shown that DM with masses around $1~\textrm{GeV}$ have a peaked distribution near the Earth's surface. While DM lighter than $1~\textrm{GeV}$ evaporate, DM heavier than $1~\textrm{GeV}$ sinks towards the Earth's center. However the presence of self interactions can arrest sinking.

\subsection{Maximum packing}

When the mediator is a vector, there are strong repulsive forces between blobs that set a limit on the local density. 
The minimum inter-blob distance $r_{\rm int}$ is given by solving 
\begin{equation}
\frac{g_{\rm blob}^2}{4\pi} \frac{e^{-m_A r_{\rm int}}}{r_{\rm int}}=T_{\rm room}
\end{equation}
To a very good approximation, this is given by,
\begin{equation}
n_{\rm max} \approx \frac{2\times10^{13}}{\textrm{cm}^3} \left(\frac{m_A}{10~\textrm{eV}}\right)^3\left(\frac{\log{10^4}}{\log{g_D}}\right)^{3.4}
\label{nmaxbml}
\end{equation}

When $m_{\rm blob} \ge 1$ GeV, evaporation is negligible \cite{Neufeld:2018slx}. In this limit, if $n_{\rm max} \lesssim n_{\rm DM}^{\rm terr}$ then, the local density anywhere on earth is set by $n_{\rm max}$. This effectively arrests sinking even for larger blob masses, since the inner regions of the Earth are saturated in blobs. If instead $n_{\rm max} \gg n_{\rm DM}^{\rm terr}$, then $n_{\rm jeans}$ sets the local number density. 

In scenario A, we will assume that an appropriate mediator mass $m_A$ can be chosen such that $n_{\rm max}\approx n_{\rm DM}^{\rm terr}$, hence the surface density is simply given by $n_{\rm surf}=n_{\rm DM}^{\rm terr}$ at all masses above $1$ GeV. Below $1$ GeV, we account for evaporation effects as treated in \cite{Neufeld:2018slx}. 

In scenario B, we will assume that the mediator mass is large enough such that $n_{\rm max} \gg n_{\rm DM}^{\rm terr}$, and assume $n_{\rm surf} =n_{\rm jeans}$ taken from \cite{Neufeld:2018slx}.

\section{Neutron bottle}
\label{sec:bottle}
The neutron bottle experiments are first summarized.\\
\begin{table}[htpb]
\centering
\begin{tabular}{|c|c|c|c|}

\hline
  Experiment &Description & trap potential [neV] & lifetime \\
  \hline
  Pattie Jr 18 \cite{Pattie:2017vsj} &grav + magnetic& 50 & $877.7\pm0.7 +0.4/-0.2$\\
   & no extrapolation & & \\ \hline
  P. Serebrov 18 \cite{serebrov2018neutron} UCN &grav +oil& $70$ & $881.5 \pm 0.7 \pm 0.6$\\
  
ARZUMANOV 15\cite{arzumanov2015measurement} & double bottle& 100&$880.2\pm 1.2$\\
STEYERL 12\cite{steyerl2012quasielastic} & material bottle & 106 &
$882.5\pm 1.4\pm 1.5$\\
PICHLMAIER 10\cite{pichlmaier2010neutron} & material bottle & 106 &
$880.7\pm 1.3\pm 1.2$ \\
SEREBROV 05\cite{serebrov2005neutron} & grav+oil trap & 106 & $878.5\pm 0.7\pm 0.3$ \\

\hline
\end{tabular}
\end{table}

We will assume that with transfer of energy larger than half of the trapping potential i.e. $E_{\rm trans}>50\textrm{neV}$ is sufficient to kick the neutrons. This sets the minimum momentum transfer, $q_{\rm min}=\sqrt{2E_{\rm trap} m_n}\approx9$eV. The cross-section to kick neutrons from the trap is given by,

\begin{equation}
  \sigma_{\rm neut}=\int dq^2 \frac{d\sigma}{dq^2} 
\end{equation}
Hence
\begin{equation}
  \sigma_{\rm neut}=\frac{4 g_{\rm blob}^2}{\pi\Lambda^2 v_{\rm th}^2}\frac{\log \left(m_A^2+q^2_{\rm max}\right)} {\log\left(m_A^2+q^2_{\rm min}\right)}\approx 10^{-34} \textrm{cm}^2 g_{\rm blob}^2 \frac{m_{\rm blob}}{\rm GeV} L_R(q_{\rm min})
   \label{sigmatheory}
\end{equation}

where the Log ratio $L_R(q_{\rm min})=\frac{\log \left(m_A^2+q^2_{\rm max}\right)} {\log\left(m_A^2+q^2_{\rm min}\right)}$ and $v_{\rm th}=\sqrt{\frac{2T_{\rm room}}{m_{\rm blob}}}$ is the thermal velocity at room temperature. Here $q_{\rm max}$ is the maximum momentum transfer set by 

\begin{equation}
q_{\rm max}=\textrm{Min}\left( R^{-1}_{\rm blob} , \mu v_{\rm th} \right)
\label{qmax}
\end{equation}

In order to explain the bottle-beam discrepancy, 
\begin{align}
  \Gamma_{\rm DM}+\Gamma_{\rm beam}=\Gamma_{\rm bottle} \nonumber \\
  n_{\rm surf} \sigma_{\rm neut}^{\rm exp} v_{\rm th} +\frac{1}{\tau_{\rm beam}}=\frac{1}{\tau_{\rm bottle}}
\end{align}

With $\tau_{\rm beam}=888 ~\textrm{sec}$, $\tau_{\rm bottle}=879 ~\textrm{sec}$. For thermalized DM, we require,
\begin{equation}
  \sigma_{\rm neut}^{\rm exp}=5.35\times10^{-25}\textrm{cm}^2 \sqrt{\frac{m_{\rm blob}}{\rm GeV}} \frac{10^{14}/\textrm{cm}^3}{n_{\rm surf}}
  \label{sigmaexp}
\end{equation}

Equating Eq.\ref{sigmaexp} with Eq.\ref{sigmatheory}, we get, 
\begin{align}
  g_{\rm blob}&\approx \frac{7.1\times 10^4}{\sqrt{L_R(9 \textrm{eV})}} \left(\frac{\rm GeV}{m_{\rm blob}}\right)^\frac{1}{4} \sqrt{\frac{10^{14}/\textrm{cm}^3}{n_{\rm surf}}}   \label{gdneutron0}
\end{align}

\begin{figure}[htpb]
\centering
\includegraphics{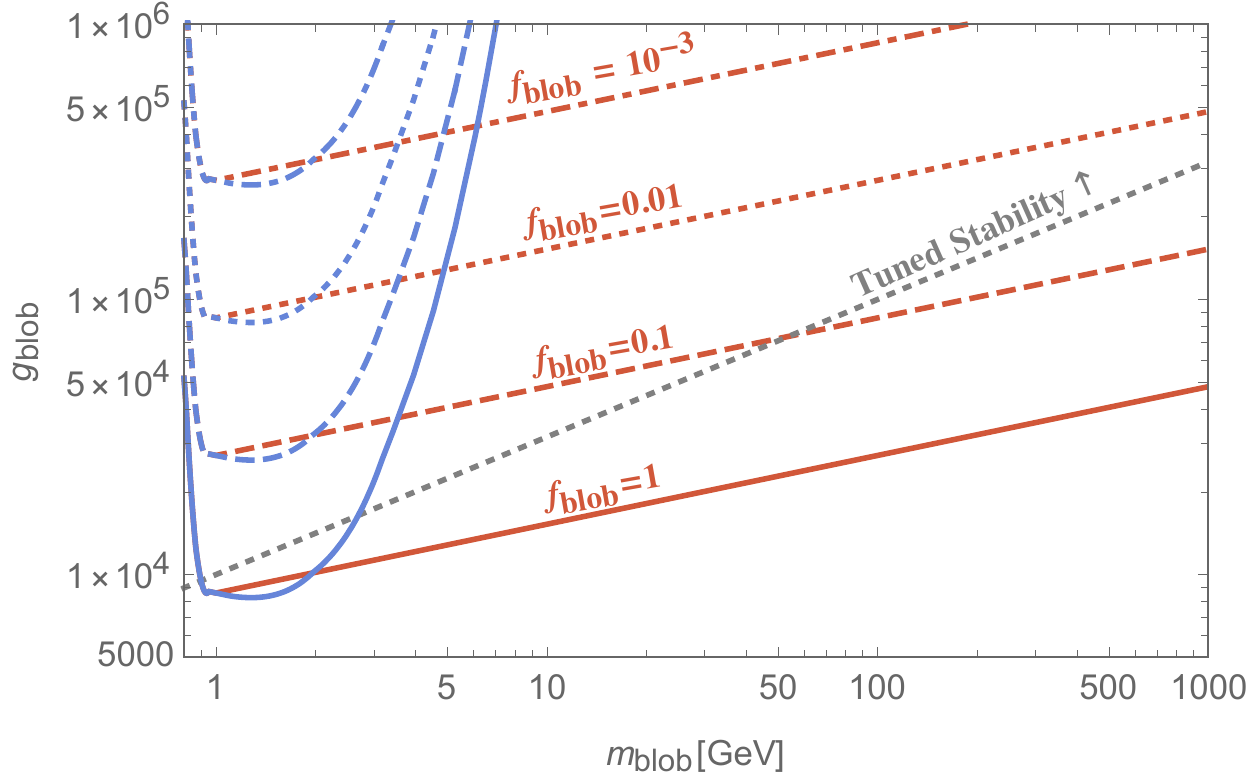}
\caption{The blob coupling $g_{\rm blob}$ that explains the observed neutron bottle anomaly is shown as a function of mass of the blob $m_{\rm blob}$ for different $f_{\rm blob}$, the fraction of dark matter in blobs. The red curves correspond to scenario A while scenario B is shown in blue. The region above the gray dashed line is tuned due to stability considerations.}

\label{fig1}
\end{figure}

This $g_{\rm blob}$ which explains the neutron lifetime anomaly, is plotted as a function of $m_{\rm blob}$ for different values of $f_{\rm blob}$, the fraction of galactic dark matter in blobs, in Fig.~\ref{fig1} approximating $L_R(9 \textrm{eV}) \approx 2$. Also shown for reference is the stability line in gray for $q_{\rm max}=R_{\rm blob}^{-1}=10$ eV for $\chi=1$ from Eqn.\ref{maxqstab}. Parameter space above this line is tuned. The red lines correspond to scenario A while blue lines correspond to scenario B. In scenario B, blobs heavier than $m_{\rm blob}\approx1$ GeV sink with small number densities at the surface, resulting in astronomically large couplings required at higher masses. In scenario A, the sinking is arrested due to blob-blob repulsions. In both scenarios, there is no viable parameter space below 1 GeV, due to significant evaporation.

We will next deal with limits from current experiments and testability at future experiments.

\section{Detection by other methods}
\label{sec:limits}
Traditional direct detection experiments, that look for virialised DM will not be sensitive to blobs considered here. This is because, the self-interactions rapidly thermalize incoming blobs, such that the blobs do not have enough kinetic energy by the time they reach even surface detectors like SENSEI~\cite{Crisler:2018gci} and CRESST~\cite{angloher2017results}. We will start by considering heating of cryogenic detectors by blobs.

\subsection{Heating of cryogenics}
\begin{figure}[htpb]
\centering
\includegraphics[width=0.8\textwidth]{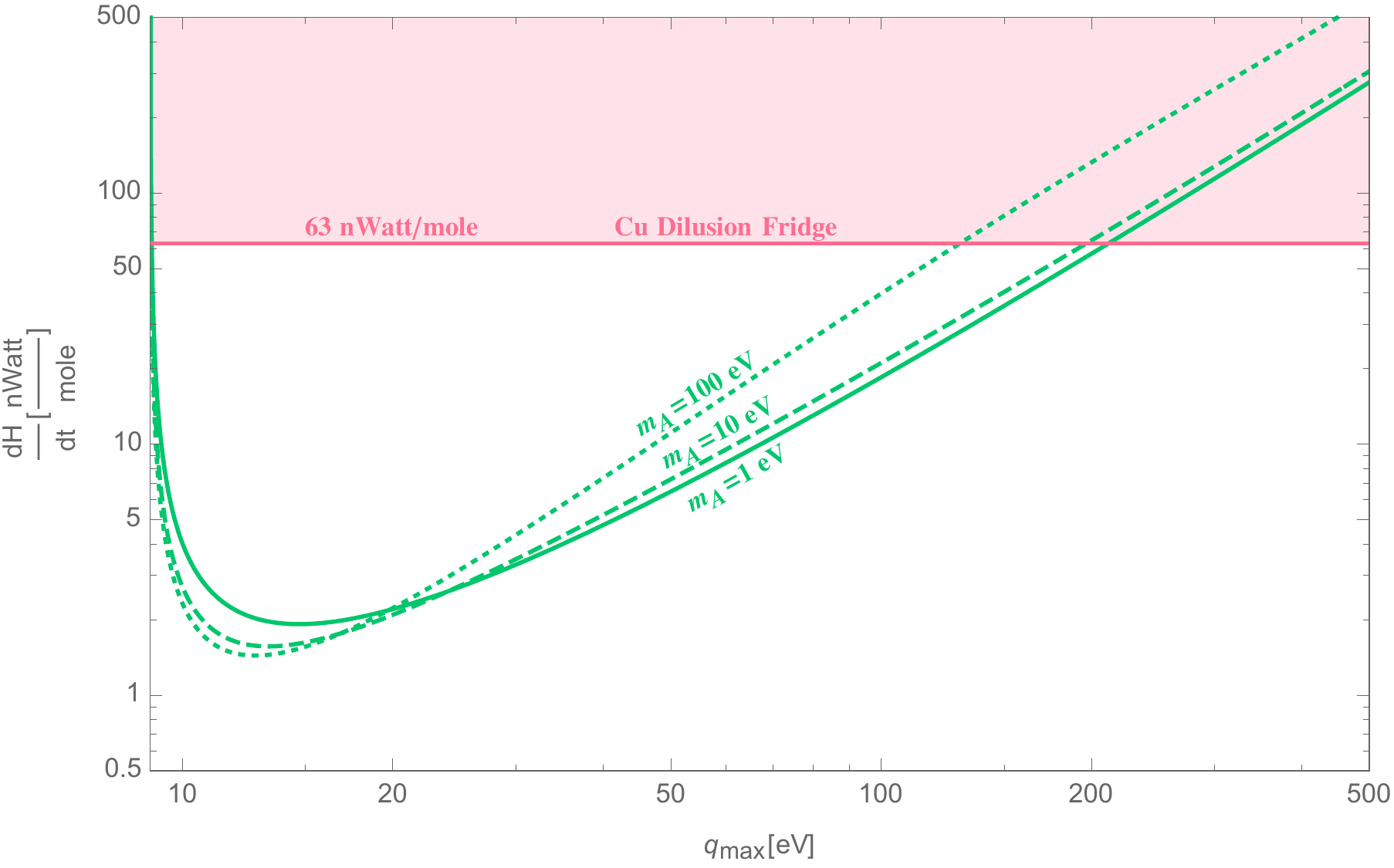}
\caption{Contours of the heat deposit rate, $\frac{dH}{dt}$ are plotted for different mediator masses $m_A$ and maximum momentum transfer $q_{\rm max}=R^{-1}_{\rm blob}$.}
\label{fig2}
\end{figure}

Unlike single event direct detection, cryogenic detectors work based on the amount of heat supplied on to the detector volume. As a result, sensitivity to smaller momentum transfers is possible. When thermalized blobs enter the detector volume, the energy averaged cross-section is given by,
\begin{equation}
 \langle\sigma E\rangle =\int \frac{d\sigma}{dq^2}\frac{q^2}{2m_T} dq^2=\frac{2 g_{\rm blob}^2}{\pi m_T v_{\rm th}^2 \Lambda^2 }\left(q_{\rm max}^2-q_{\rm min}^2 +m_A^2 L_R(0)\right)
\end{equation}
here $m_T$ is the mass of the target atom. 
The energy deposition rate per target atom is,
\begin{equation}
\frac{dH}{dt} = \langle\sigma E\rangle n_{\rm DM} v_{\rm th}
\end{equation}
Substituting for $g_{\rm blob}$ from Eqn.~\ref{gdneutron0}, the heating rate caused by DM that can explain the neutron bottle anomaly is,

\begin{equation}
\frac{dH}{dt}\left(anomaly\right) = \frac{56}{A_T} \frac{\textrm{nWatt}}{\rm mole} \left(\frac{q_{\rm max}^2-q_{\rm min}^2 +m_A^2 L_R(0)}{(10 \textrm{eV})^2 L_R(9 \textrm{eV})}\right)
\label{target}
\end{equation}
This is $m_A$ and $q_{\rm max}$ dependent. The heat deposit rate $\frac{dH}{dt}$ due to blobs that explain the neutron bottle anomaly is plotted in Fig.~\ref{fig2} in green as a function of $q_{\rm max}$ described in Eqn.~\ref{qmax} for different $m_A$. $A_T=63$ corresponds to copper. For this range $q_{\rm max} = R^{-1}_{\rm blob}$. 

The heating rate of several cryogenic compounds was investigated in Refs.~\cite{Neufeld:2018slx} and \cite{Neufeld:2019xes}. In Ref.~\cite{Neufeld:2018slx}, the heating rate of Helium, Hydrogen, Nitrogen, Oxygen and Argon were reported as a fractional evaporation $\mathscr{P}/\textrm{day}$. This can be converted into heating rate per mole through,
\begin{equation}
\frac{dH}{dt}=\frac{\mathscr{P}}{\textrm{day}} L\times A_T \approx 10 \frac{\mu \textrm{Watt}}{\rm mole} \frac{\mathscr{P}}{0.01} \frac{L}{26 \textrm{Joule}/\textrm{gram}} \frac{A_T}{4}
\end{equation}
Here $L$ is the latent heat of evaporation. The above equation is normalized to Helium and is an order of magnitude larger than Eqn.(\ref{target}) and hence does not constrain. As seen in Table 1 of Ref~\cite{Neufeld:2018slx}, the other elements, Hydrogen, Nitrogen, Oxygen and Argon have much larger latent heat and hence are less constraining. It is also important to remember that the above estimate was made in the transparent dewar limit, and for large enough cross-sections, blobs can thermalize with the outside of the flask and cool, thereby causing lesser heating of the cryogen. Table 1 of \cite{Neufeld:2019xes} also lists heating rate in $\mu \textrm{Watt}/{\rm mole}$. All their limits are larger than $10 \mu \textrm{Watt}/{{\rm mole}}$. We find that these limits are sub-leading compared to ones from dilution fridges. 

The stability of dilution fridges could also set limits on the heating rate. Roughly we take this to be $100\frac{\textrm{nWatt}}{\rm mole}$ in a 100 gram sample of copper \cite{private}. This translates to $\frac{dH}{dt}=63\frac{\textrm{nWatt}}{\rm mole}$ which is plotted in pink in Fig.~\ref{fig2}.
The region below this Pink curve and above a green curve explains the anomaly and is not constrained yet. All of the parameter space could be tested with sensitivity to a heat deposit rate of about $1\frac{\textrm{nWatt}}{\rm mole}$.

\subsection{High momentum probes}

The orbital decay of HST was used to set limits on strongly interacting models in \cite{Neufeld:2018slx}. Roughly the constraint corresponded to,
\begin{align}
\frac{dR}{dt}&= 2 n \sigma_T \frac{\mu}{m_{\rm HST}} R_{\rm orb} v_{\rm orb}
\end{align}
This primarily probes momentum transfers $\mu v_{\rm orb}\sim 100 \textrm{keV}$, orders of magnitude larger than the inverse size of the blobs we consider. For this reason, blobs which explain the neutron bottle anomaly predict $\frac{dR}{dt}\sim 1\frac{\textrm{m}}{\rm year}$, while the limit currently is a few orders of magnitude higher at $0.8 \frac{\rm km}{yr}$, 

Metastable Tantalum was proposed as a probe of dark matter accumulated on earth in \cite{Pospelov:2019vuf,Lehnert:2019tuw}. However, the momentum transfer required for de-excitation of Tantalum $q_0 \approx 1 \textrm{MeV}$ suppresses the long range force tremendously,
  \begin{equation}
  \sigma_{\rm Ta}=\frac{\mu g_{\rm SM}^2 g_{\rm blob}^2}{2\pi q_0^3} S(q_0)\approx 10^{-49} \textrm{cm}^2 \sqrt{\frac{\rm GeV}{m_{\rm blob}}} g_{\rm blob}^2 S(q_0)
\end{equation}
where $q_0\approx 12.5$keV. Note that $S(q_0)$ is suppressed severely as well for $q_0\ll 1$MeV. As a result these limits are not relevant either.

The regime relevant to limits from LHC beam lifetime are around 6.5 TeV \cite{Neufeld:2018slx}. Only inelastic collisions affect the beam as the multipoles correct for small changes in momentum. As a result this is not expected to constrain these blobs either. 

Lastly, anomalous heat transport on earth and in the sun, have been used to set limits on exotic matter that has collected over time \cite{Mack:2007xj}. However, the self-interaction cross-section is so high that thermal conductivity is very low. As a result these do not set appreciable limits on blobs with large coupling.

\subsection{Fifth Force experiments}
Experiments looking for fifth forces typically operate by measuring the displacement of a test mass mass due to a force felt by another mass. With 1 sec (3000 sec) sensitivity, these experiments are sensitive to $0.2$ nm displacements with interferometry technique. We estimate here, the displacement caused by the thermal distribution of dark matter blobs on a $5\times 10^{-5}$ gram test mass. 

For a blob that traverses a distance $b$ from the mass, the force is given by,
\begin{equation}
F_{\rm blob} = \sqrt{N_{\rm spin}} \frac{g_{\rm blob}}{\Lambda b^3} 
\end{equation} 
 And hence the resultant displacement is given by,
 \begin{equation}
 d(b)= \sqrt{N_{\rm spin}} \frac{g_{\rm blob}}{m_{\rm test} \Lambda b^3} \left(\frac{b}{v}\right)^2
 \end{equation}
 The resultant net displacement due to several blobs traversing is obtained by an incoherent sum of these distances and hence given by, 
 \begin{equation}
 D^2=\int d^2(b) f(b) db = N_{\rm spin}\frac{t}{v^3}\left( \frac{g_{\rm blob}}{m_{\rm test} \Lambda }\right)^2 \log\left(\frac{b_{\rm max}}{b_{\rm min}}\right)
 \end{equation}
 Here $f(b)=b n_{\rm blob} v t$, $b_{\rm max}= \frac{1}{m_A}, b_{\rm min}=R_{\rm blob}$,
 This is approximately,
 \begin{equation}
 D= 10^{-23} \left( \frac{m_{\rm blob} }{\rm GeV} \right)^\frac{3}{4} \left(\frac{n_{\rm blob}}{10^{14} /\textrm{cm}^3}\right)^{\frac{1}{2}}\textrm{meter}.
 \end{equation}
 This is orders of magnitude smaller than the sensitivity of interferometers. 
 
\section{Conclusions}
\label{sec:conclusions}
We have shown that the anomalously small lifetime of the neutron in trapped neutron experiments can be explained by the soft scattering of a bound population of $\approx$ GeV scale dark matter with neutrons. This phenomenology is only possible with composite dark matter where soft scattering can have a considerably larger cross-section than hard scattering events which are otherwise constrained by direct detection experiments. 

This phenomenology motivates two kinds of experimental strategies. First, trapped neutron experiments could check our proposed solution by placing more neutron detectors around the trap covering a larger solid angle, thus observing the scattered neutrons. It would also be interesting to see if the measured lifetime changes by using a stronger magnetic field to confine the neutrons in the trap. The magnetic trap could potentially be made a factor of 10x stronger. One might suspect that dark matter models could be constructed to deposit this larger amount of energy. While there is some freedom to do this, the parameter space is constrained by the analysis shown in Fig.~\ref{fig1}. 

Second, irrespective of the anomaly in the lifetime of the neutron, we have identified a new experimental opportunity in dark matter detection. Composite dark matter naturally gives rise to enhanced soft scattering while having a highly reduced hard scattering cross-section. This can lead to a cold population of dark matter with a significantly enhanced number density that is bound to the Earth. This population cannot be detected using conventional dark matter detectors that are aimed at identifying the ionization and scintillation produced as a result of hard collisions. But, the total energy deposited by the dark matter in these soft collisions is quite significant and it may be possible to detect it using a dedicated setup~\cite{forthcoming}. 

\begin{acknowledgments}
We thank Julien Billard, Matt Pyle and Andy Saunders for useful discussions. S.R. is supported by the US National Science Foundation (contract No. PHY-1818899). S.R. is also supported by the DoE under a QuantISED grant for MAGIS. H.R. is supported in part by the DOE under contract DE-AC02-05CH11231.
\end{acknowledgments}

\bibliographystyle{unsrt}
\bibliography{biblio}
\appendix
\section{Capture}
\label{seccap}
Dark matter blobs falling on Earth will get captured if they slows down to velocities below the escape velocity. To estimate the slow down, we first present the momentum transfer cross-section. 

\begin{equation}
  \sigma_T(v) = \frac{g_{\rm blob}^2 q_{\rm max}^2}{8\pi\mu^2 \Lambda^2 v^4}
  \label{transfer}
\end{equation}

For initial capture, we require
\begin{equation}
  \sigma_{\rm T}(v)\gtrsim \frac{m_{\rm blob}}{\mu n_{\rm rock} R_{\rm Earth}}
\end{equation}
The coupling corresponding to capturing all of DM incoming with velocity below v corresponding to,
\begin{equation}
  g_{\rm blob}^{\rm cap}(v) \approx 10^{15} v^2 \frac{10 \textrm{eV}}{q_{\rm max}} \frac{\sqrt{\mu \times m_{\rm blob}}}{\rm GeV}
\end{equation}

For $g_{\rm blob} \ge  g_{\rm blob}^{\rm cap}(10^{-3})$, $\mathcal{O}(1)$ of incoming blobs is captured. For $g_{\rm blob} < g_{\rm blob}^{\rm cap}(10^{-3})$, only a slow moving fraction $ f_{\rm cap}\propto v^3$ is captured. This leads to a build-up of dark matter on earth, this is given by,

\begin{equation}
 \langle R_{\rm DM}^{\rm terr} \rangle =f_{\rm cap}(g_{\rm blob}) \frac{\pi r_E^2}{\frac{4}{3}\pi r_E^3} n_{\rm DM}^{\rm vir} v_{\rm vir} \approx\frac{10^{5}}{{\rm cm}^3} \frac{1}{\rm year} f_{\rm cap}(g_{\rm blob}) f_{\rm blob} \frac{\rm GeV}{m_{\rm blob}}
 \label{capturest}
\end{equation}

\subsection{Self-interactions}

After initial accumulation of DM, self-interactions start becoming important. For blob-blob scattering, the Born approximation is not valid for the large couplings we consider and we work in the classical regime. The self-interaction cross-section is given by,

\begin{align}
  \sigma_{\rm self}\approx \frac{\pi }{m_A^2} \approx 10^{-9}\textrm{cm}^2 \left(\frac{\rm eV}{m_A}\right)^2 
\end{align}

Thus the self-interactions dominate capture dynamics when
\begin{equation}
n_{\rm blob} \sigma_{\rm self} \ge n_{\rm rock} \sigma_{\rm T}
\label{selfdomi}
\end{equation}

We will assume that all self-interactions result in an incoming blob with velocity $v_{\rm vir}$ transferring half its momentum into thermalized blobs. Hence, incoming and thermalized blobs both leave with velocity $\frac{v_{\rm vir}}{2}$ due to a single collision. There is subsequent capture only if the penetration depth of these blob particles is smaller than the depth of the remaining rock or terrestrial dark matter population. Otherwise, this results in net loss of dark matter on earth. For large enough coupling this is a runaway process and there is $100\%$ capture of virial blobs due to subsequent self interactions. The exact calculation of this coupling involves multiple levels of blob-blob collisions and will require a Monte Carlo simulation which we leave for future work. However, considering up to two collisions, we have checked that $g_{\rm blob} \gtrsim \frac{1}{4} g_{\rm blob}^{\rm cap}(10^{-3})$ is enough to capture close to $100\%$ of all blobs falling in. 

Whereas for smaller coupling this leads to negligible build-up due to incoming blobs kicking thermalized blobs out. Number densities are expected to saturate the equality in Eqn.~\ref{selfdomi}, i.e.
\begin{equation}
n_{\rm eq}^{\rm kick} \sim n_{\rm rock} \frac{\sigma_{\rm T}}{\sigma_{\rm self}}\approx \frac{1}{{\rm cm}^3}
\end{equation}

Blobs with small couplings can be captured if there is a secondary population of more massive blobs thermalized on earth.  
 \subsection{Secondary blobs}
We assume that there is a secondary blob population of type $B$ with mass $m_{\rm blob}^B$ with charge $g_{\rm blob}^B$, with $\frac{m_{\rm blob}^A}{g_{\rm blob}^A}= \frac{m_{\rm blob}^B}{g_{\rm blob}^B}$. Here $A$ type blobs are the ones populating the earth's surface and responsible for the neutron bottle anomaly. We further assume that $B$ blobs have larger mass (and charge) i.e. $m_{\rm blob}^B \gg m_{\rm blob}^A$ but make up a much smaller fraction of DM, $f_{\rm blob}^B \ll f_{\rm blob}^A$. This is a natural consequence of dark nucleosynthesis. We will assume the secondary population satisfies $g_{\rm blob}^B > g_{\rm blob}^{\rm cap}(10^{-3})$ such that all of its in-falling virial population gets captured. While the incoming $A$ blobs with virial velocity continue to kick the thermalized population of $A$ blobs, they do not carry enough momentum to kick out blobs of type $B$. Since the self-interaction cross-sections are identical for both species, kick-out due to $A-A$ self interactions become important only when $n_A \ge n_B$. However if 
\begin{align} 
n_B \sigma_{\rm self} R_E &\gg 1 \nonumber \\
n_B &\gg \frac{1.5}{\rm cm^3} \left(\frac{m_A}{\rm eV}\right)^2 
\end{align}
 then by the time the $A$ blobs build up to $n_B$ density, there is enough of terrestrial $A$ blobs to absorb all of the momentum from incoming $A$ blobs and prevent subsequent kick-out. Then subsequent self-interactions rapidly build-up $A$ blobs; a runaway process which can soon outnumber $B$ blobs due to their superior virial density. Due its large mass, heavy $B$ blobs subsequently sink to the Earth's core and are a non-factor in experiments today. Thus an extremely small number density of a heavier population of blobs can help achieve a near $100\%$ capture of $A$ blobs on earth. 

\end{document}